# Higher order closed-form model for isotropic hyperelastic spherical shell (3D solid).

E Hanukah

Faculty of Mechanical Engineering, Technion – Israel Institute of Technology, Haifa 32000, Israel

Email: eliezerh@tx.technion.ac.il

**Abstract**

In the present study we extend the finite body formulation to isotropic compressible Neo-Hookean shell, namely, we derive a set of ordinary differential equations (ODE) for the dynamics of a spherical shell with general 3D kinematics. Each ODE explicitly depends on material geometrical and load parameters. The number of the governing equations matches the number of internal degrees of freedom (IDF), which is controlled by the desired order of the model. The first order model allows constant deformation gradient, accounts for homogeneous deformation, and includes twelve IDF, while a the second order enables linear variation of deformation gradient in the domain, resulting in thirty IDF, and so on... First to fifth order models have been considered in the study. Initial shell configuration is represented in an exact manner, while systematic kinematic approximation, equivalent to Taylor multivariate expansion, is used to express the actual configuration in terms of IDFs. No symmetries or special deformation modes were invoked, enabling the model to capture general-purpose 3D phenomena. Later on, coordinates transformation imposed trigonometric terms to simple algebraic shape functions. Standard weak (Bubnov-Galerkin) formulation was adopted, making incorporation of existing constitutive equations possible. Compressible Neo-Hookean material law is expressed using approximated kinematics and then systematically approximated, such that analytic integration for internal forces is easily performed, making numerical integration unnecessary. No numerical or closed-form study of nonlinear problem is presented herein, however, the linearized formulation is applied to small vibration problem, convergence is demonstrated and a new simple explicit expression for fundamental frequency of free vibrating hemisphere is found.

## 1. Introduction

Growing computer power enables engineers to develop and use general formulations for curved surfaces, without assuming any approximations (e.g. the size of rotations, deflections, cross section behavior etc. e.g. [1-4])

The presented formulation is ideologically similar to p-version of FEM, nevertheless, details of kinematic approximation here is different, for example, kinematic approximation given in terms of IDF and not in terms of nodes. Moreover, first Piola-Kirchhoff stress tensor is systematically approximated, such that closed-form integration is performed, numerical integration points are not used (e.g. closed-form Neo-Hookean based formulation [5], for linear elasticity [6-9]).

In the last several decades there has been an effort to provide 3D elasticity solutions for the free vibration of prisms, parallelepipeds, cylinders etc., vast majority of these works have utilized various techniques to obtain *numerical results* (e.g. [10-27]), and for some special cases, closed-form results were achieved (e.g. [28, 29]) by the use of Cosserat Point theory and Pseudo Rigid Body approach, however, these approaches, to the best of our knowledge, do not have systematic generalization to higher orders (e.g. [30, 31])

The present study follows basic guidelines given in [5, 32-34] and excels it to the curved spherical surface. Similarly to finite element method (e.g.[35, 36]), especially its p-version (e.g. [37]), the formulation has two cornerstones; kinematic approximation and weak (Bubnov-Galerkin) formulation (or variational formulation for FEM).

The outline of the paper is as follows. Section 2 presents the main theoretical considerations for deriving the non-linear dynamical equations of motion. Governing equations are obtained by adopting a weak formulation combined with analytical integration. Linearization of the governing equations with respect to the internal degrees of freedom is discussed in Section 3. Assuming small vibrations the mass and stiffness matrices are defined and an eigenvalue problem for the natural frequencies and modes is formulated. In Section 4, the solutions of free vibration problem are listed and discussed. The accuracy of closed-form expressions for the natural frequencies is examined by comparison to finite-elements simulations.

## 2. Theoretical considerations

Consider a three dimensional spherical shell, see Fig. 1, occupying a finite volume in Euclidean space, made of isotropic and hyperelastic (Neo-Hookean) material. First, the basic equations of elasticity are recalled. (1) is the balance of linear momentum in initial configuration. Balance of angular momentum implies (2), where $\mathbf{P}$ is the first Piola-Kirchhoff stress tensor, $\mathbf{F}$ is deformation gradient $\mathbf{F} = \text{Grad}(\mathbf{x})$, where gradient operator is defined by $\text{Grad}(\bullet) = \sum_{k=1}^{3} (\bullet)_{,k} \otimes \mathbf{G}^k$ and comma stand for partial differentiation with respect to coordinates, $\otimes$ stand for tensor/outer product, right transpose of arbitrary second order tensor $\mathbf{A}$ is denoted by $\mathbf{A}^T$. Constitutive Neo-Hookean stress-strain relations (material law) is given by (3) (e.g. [35] pp. 45):

$$\rho_0 \dot{\mathbf{v}} = \rho_0 \mathbf{b} + \text{Div}(\mathbf{P}) \qquad (1)$$

$$\mathbf{PF}^T = \mathbf{FP}^T \qquad (2)$$

$$\mathbf{P} = \left(\frac{\lambda}{2}(J^2 - 1)\mathbf{I} + \mu(\mathbf{FF}^T - \mathbf{I})\right)\mathbf{F}^{-T} \qquad (3)$$

Above, $\mathbf{X}$ and $\mathbf{x}$ represent the locations of a material point X in the initial and actual configurations, respectively, J denotes determinant of deformation gradient $J = \det(\mathbf{F})$, $\rho_0$ denotes initial mass density, $\lambda$ and $\mu$ are the Lame constants, $\mathbf{v} = \dot{\mathbf{x}}$ is the velocity of a material particle, a superposed dot denotes time differentiation, $\mathbf{b}$ stands for body force per unit of mass, divergence operator is given by $\text{Div}(\bullet) = \sum_{k=1}^{3}(\bullet)_{,k}\, \mathbf{G}^k$, inverse of arbitrary invertible second order tensor $\mathbf{A}$ is denoted by $\mathbf{A}^{-1}$, $\mathbf{I}$ stands for second order identity tensor. Standard relations between material constants are recalled $\lambda = E\nu / ((1+\nu)(1-2\nu))$, $\mu = E/2(1+\nu)$, where E and $\nu$ are the Young's modulus and Poisson's ratio. Restriction (2) is identically satisfied by (3). Standard transformation between Cartesian and spherical coordinates is recalled (see Fig. 1)

$$\hat{X}_1 = r\cos(\theta)\sin(\phi)\ ,\ \hat{X}_2 = r\sin(\theta)\sin(\phi)\ ,\ \hat{X}_3 = r\cos(\phi)\ ,\ dV = r^2\sin(\phi)drd\phi d\theta \qquad (4)$$

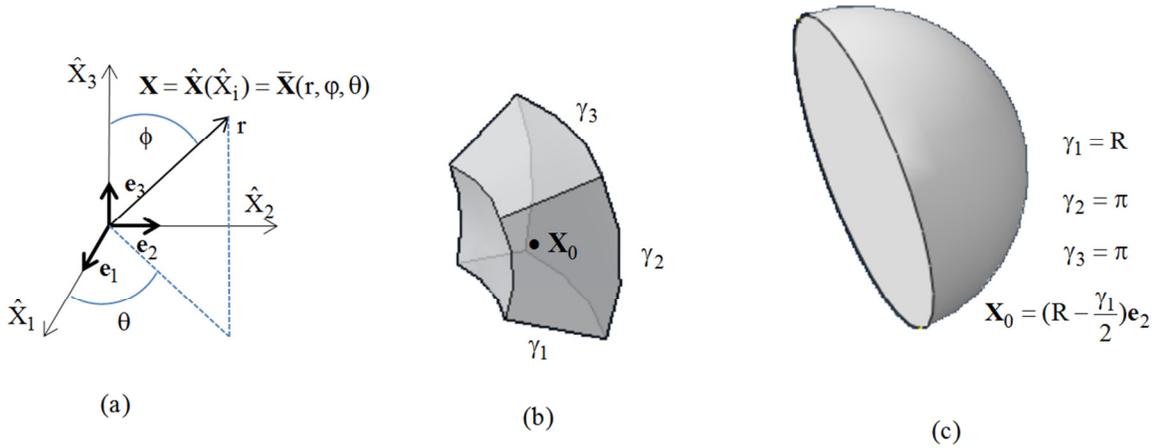

Figure 1: Schematic illustration showing the 3D spherical shell in its initial configuration. (a) showing the local Cartesian orthonormal base vectors $\{\mathbf{e}_1,\mathbf{e}_2,\mathbf{e}_3\}$, their material coordinates $\{\hat{X}_1,\hat{X}_2,\hat{X}_3\}$, cylindrical material coordinates $\{r,\phi,\theta\}$. (b) showing initial configuration of thick spherical shell, its thickness controlled by $\gamma_1$ while $\gamma_2, \gamma_3$ controls range in $\phi, \theta$ respectively, $\mathbf{X}_0$ is the center point. (c) showing specific choice of parameters such that in the initial configuration shell becomes half a sphere.

## 2.1 Kinematic approximation.

The middle point of initial configuration of a shell denoted by $\mathbf{X}_0 = \sum_{k=1}^{3} \hat{X}_{0k}\mathbf{e}_k$, (see Fig. 1). Cartesian base vectors $\{\mathbf{e}_1, \mathbf{e}_2, \mathbf{e}_3\}$ and corresponding material coordinates $\{\hat{X}_1, \hat{X}_2, \hat{X}_3\}$ are used to represent the material point $X \in \Omega_0$ occupying position $\mathbf{X}$ in initial configuration. For later convenience, local coordinate systems (Cartesian and spherical) are chosen such that the middle point is given by

$$\mathbf{X}_0 = \hat{X}_{01}\mathbf{e}_1 + \hat{X}_{02}\mathbf{e}_2 + \hat{X}_{03}\mathbf{e}_3 = 0\mathbf{e}_1 + (R - \frac{\gamma_1}{2})\mathbf{e}_2 + 0\mathbf{e}_3 \tag{5}$$

$$\mathbf{X} = \hat{X}_1\mathbf{e}_1 + \hat{X}_2\mathbf{e}_2 + \hat{X}_3\mathbf{e}_3 = \mathbf{X}_0 + (\hat{X}_1 - \hat{X}_{01})\mathbf{e}_1 + (\hat{X}_2 - \hat{X}_{02})\mathbf{e}_2 + (\hat{X}_3 - \hat{X}_{03})\mathbf{e}_3$$

Where R is the external radius and $\gamma_1$ is the width of the shell (see Fig. 1). Following the above, $\gamma_i$ (i = 1, 2, 3) define the domain by

$$R - \gamma_1 \leq r \leq R \ , \ \frac{\pi}{2} - \frac{\gamma_2}{2} \leq \phi \leq \frac{\pi}{2} + \frac{\gamma_2}{2} \ , \ \frac{\pi}{2} - \frac{\gamma_3}{2} \leq \theta \leq \frac{\pi}{2} + \frac{\gamma_3}{2} \tag{6}$$

Using the shape functions definition

$$N_0 = 1 \ , \ N_1 = \hat{X}_1 - \hat{X}_{01} \ , \ N_2 = \hat{X}_2 - \hat{X}_{02} \ , \ N_3 = \hat{X}_3 - \hat{X}_{03}$$
$$N_4 = (\hat{X}_1 - \hat{X}_{01})(\hat{X}_2 - \hat{X}_{02}) \ , \ N_5 = (\hat{X}_1 - \hat{X}_{01})(\hat{X}_3 - \hat{X}_{03}) \ , \ N_6 = (\hat{X}_2 - \hat{X}_{02})(\hat{X}_3 - \hat{X}_{03})$$
$$N_7 = (\hat{X}_1 - \hat{X}_{01})(\hat{X}_1 - \hat{X}_{01}) \ , \ N_8 = (\hat{X}_2 - \hat{X}_{02})(\hat{X}_2 - \hat{X}_{02}) \ , \ N_9 = (\hat{X}_3 - \hat{X}_{03})(\hat{X}_3 - \hat{X}_{03})... \tag{7}$$
$$N_i = \hat{N}_i(\hat{X}_1, \hat{X}_2, \hat{X}_3) = \bar{N}_i(r, \phi, \theta) \ , \ (i = 0, .., n^{shf})$$

One can rewrite representation (5) as

$$\mathbf{X} = \sum_{j=0}^{n^{shf}} N_j \mathbf{X}_j \ , \ \mathbf{X}_j = \mathbf{e}_j \ (j = 1, 2, 3) \ , \ \mathbf{X}_j\big|_{j>3} = 0 \ , \ n^{shf} \geq 3 \tag{8}$$

Using Lagrangian description, the actual configuration $\mathbf{x}$ of a material point X is given by the motion $\mathbf{x} = \chi(\hat{\mathbf{X}}(\hat{X}_1, \hat{X}_2, \hat{X}_3), t)$ or equivalently $\mathbf{x} = \chi(\bar{\mathbf{X}}(r, \phi, \theta), t)$. The mapping function $\chi$ is unknown, and has to be determined. Finding the exact mapping is usually not possible therefore some approximations have to be considered. Here, we systematically approximate the motion function $\chi$ by means of Taylor's multivariable expansion about the middle point of the body $\mathbf{X}_0 = \chi(\hat{\mathbf{X}}(0,0,0), 0)$ with respect to Cartesian coordinates. To this end we define the next operator

$$(\bullet)_0 = (\bullet)|_{\mathbf{X}=\mathbf{X}_0} \ , \ (\bullet)_1 = \frac{\partial(\bullet)}{\partial \hat{X}_1}\bigg|_{\mathbf{X}=\mathbf{X}_0} \ , \ (\bullet)_2 = \frac{\partial(\bullet)}{\partial \hat{X}_2}\bigg|_{\mathbf{X}=\mathbf{X}_0} \ , \ (\bullet)_3 = \frac{\partial(\bullet)}{\partial \hat{X}_3}\bigg|_{\mathbf{X}=\mathbf{X}_0}$$

$$(\bullet)_4 = \frac{\partial^2(\bullet)}{\partial \hat{X}_1 \partial \hat{X}_2}\bigg|_{\mathbf{X}=\mathbf{X}_0} \ , \ (\bullet)_5 = \frac{\partial^2(\bullet)}{\partial \hat{X}_1 \partial \hat{X}_3}\bigg|_{\mathbf{X}=\mathbf{X}_0} \ , \ (\bullet)_6 = \frac{\partial^2(\bullet)}{\partial \hat{X}_2 \partial \hat{X}_3}\bigg|_{\mathbf{X}=\mathbf{X}_0} \quad (9)$$

$$(\bullet)_7 = \frac{\partial^2(\bullet)}{2\partial \hat{X}_1 \partial \hat{X}_1}\bigg|_{\mathbf{X}=\mathbf{X}_0} \ , \ (\bullet)_8 = \frac{\partial^2(\bullet)}{2\partial \hat{X}_2 \partial \hat{X}_2}\bigg|_{\mathbf{X}=\mathbf{X}_0} \ , \ (\bullet)_9 = \frac{\partial^2(\bullet)}{2\partial \hat{X}_3 \partial \hat{X}_3}\bigg|_{\mathbf{X}=\mathbf{X}_0} \ .....$$

Using definition (7),(9) and defining $\mathbf{x}_j = \chi_j, (j=0,..,n^{shf})$, Taylor's multivariable expansion of $\mathbf{x}$ is given by

$$\mathbf{x}^h = \sum_{j=0}^{n^{shf}} N_j \mathbf{x}_j(t)$$

$$\alpha = 1 \Rightarrow n^{shf} = 3 \ , \ \alpha = 2 \Rightarrow n^{shf} = 9 \ , \ \alpha = 3 \Rightarrow n^{shf} = 19 \quad (10)$$

$$\alpha = 4 \Rightarrow n^{shf} = 34 \ , \ \alpha = 5 \Rightarrow n^{shf} = 55 \quad ........$$

Here and throughout the text, upper symbol $()^h$ denotes approximated function so that $\mathbf{x}^h$ stands for approximation of $\mathbf{x}$. Here, the shape functions (7) and the partial derivatives (9) are explicitly stated up to second order, however it is emphasized, that the shape functions and the partial derivatives follow from multivariable expansion about the middle point. Order of approximation is denoted by $\alpha$. Partial derivatives of $\chi$ are evaluated at the middle point, as a result they are merely functions of time and unknown. Separation of variables is achieved. The order of approximation specifies the number of terms in the above approximation.

To determine the initial values of $\mathbf{x}_j(t)$, it is recalled that approximation (10) has to admit (8) for the initial configuration, so the initial values of unknown functions $\mathbf{x}_j(t)$ are

$$\mathbf{x}^h\bigg|_{\substack{t\to 0 \\ \chi\to \mathbf{1}}} = \mathbf{X} \ \Rightarrow \ \mathbf{x}_j\bigg|_{\substack{t\to 0 \\ \chi\to \mathbf{1}}} = \mathbf{X}_j \ , \ (j=0,..,n^{shf}) \quad (11)$$

where $\mathbf{1}$ stands for identity transformation. Next, displacement field becomes

$$\mathbf{u}^h = \mathbf{x}^h - \mathbf{X} = \sum_{j=0}^{n^{shf}} N_j \mathbf{u}_j(t)$$

$$\mathbf{u}_j(t) = \mathbf{x}_j(t) - \mathbf{X}_j \ , \ \mathbf{u}_j\bigg|_{\substack{t\to 0 \\ \chi\to \mathbf{1}}} = \mathbf{0} \ , \ (j=0,..,n^{shf}) \quad (12)$$

Since (12) implies that all $\mathbf{u}_j(t), (j=0,..,n^{shf})$ are zero in the initial configuration, it is natural to define the internal degrees of freedom (IDF) as their components, i.e.

$$b_{3j+k}(t) = \mathbf{u}_j(t) \cdot \mathbf{e}_k \quad , \quad \mathbf{u}_j(t) = \sum_{k=1}^{3} b_{3j+k}(t)\mathbf{e}_k$$

$$b_m(t)\Big|_{\substack{t \to 0 \\ \chi \to 1}} = 0 \quad , \quad (j=0,..,n^{shf}, m=1,..,n^{dof}) \tag{13}$$

$$n^{dof} = 3(n^{shf}+1) \tag{14}$$

Standard definitions of covariant and contravariant base vectors in initial configuration and some relations between them are recalled

$$\mathbf{G}_k = \partial \mathbf{X}/\partial \hat{X}_k = \mathbf{X}_{,k} = \mathbf{e}_k \ , \ (k=1,2,3) \ , \ G^{1/2} = \mathbf{G}_1 \times \mathbf{G}_2 \cdot \mathbf{G}_3 = 1 > 0 \tag{15}$$

$$\mathbf{G}^1 = \frac{\mathbf{G}_2 \times \mathbf{G}_3}{G^{1/2}} \ , \ \mathbf{G}^2 = \frac{\mathbf{G}_3 \times \mathbf{G}_1}{G^{1/2}} \ , \ \mathbf{G}^3 = \frac{\mathbf{G}_1 \times \mathbf{G}_2}{G^{1/2}} \tag{16}$$

Where $\delta_m^n$ is the Kronecker's delta, $(\cdot)$ stands for scalar product and $(\times)$ for vector product, comma $(\cdot)_{,i}$ $(i=1,2,3)$ denotes partial differentiation with respect to $\hat{X}_k$. Next we proceed to the following standard definitions of covariant and contravariant bases and their relations

$$\mathbf{g}_k = \partial \mathbf{x}^h / \partial \hat{X}_k = \mathbf{x}^h{}_{,k} \ , \ g^{1/2} = \mathbf{g}_1 \times \mathbf{g}_2 \cdot \mathbf{g}_3 > 0$$

$$\mathbf{g}^1 = \frac{\mathbf{g}_2 \times \mathbf{g}_3}{g^{1/2}} \ , \ \mathbf{g}^2 = \frac{\mathbf{g}_3 \times \mathbf{g}_1}{g^{1/2}} \ , \ \mathbf{g}^3 = \frac{\mathbf{g}_1 \times \mathbf{g}_2}{g^{1/2}} \ , \ \mathbf{g}_k \cdot \mathbf{g}^m = \delta_k^m \ , \ (k,m=1,2,3) \tag{17}$$

where comma denotes partial differentiation. Also, using the definitions of the deformation gradient, $\mathbf{F} = \dfrac{\partial \mathbf{x}}{\partial \mathbf{X}}$, together with (15),(16) and (17), the approximated forms of the kinematic tensors are

$$\mathbf{F}^h = \sum_{k=1}^{3} \mathbf{g}_k \otimes \mathbf{G}^k \quad , \quad (\mathbf{F}^h)^T = \sum_{k=1}^{3} \mathbf{G}^k \otimes \mathbf{g}_k ,$$

$$(\mathbf{F}^h)^{-1} = \sum_{k=1}^{3} \mathbf{G}_k \otimes \mathbf{g}^k \ , \ (\mathbf{F}^h)^{-T} = \sum_{k=1}^{3} \mathbf{g}^k \otimes \mathbf{G}_k \tag{18}$$

$$\tilde{\mathbf{F}}^{-T} = J^h (\mathbf{F}^h)^{-T} = \frac{1}{G^{1/2}} \sum_{k=1}^{3} \tilde{\mathbf{g}}^k \otimes \mathbf{G}_k \ , \ \tilde{\mathbf{g}}^m = g^{1/2} \mathbf{g}^m \ (m=1,2,3) \tag{19}$$

Implementing the above in Neo-Hookean (3) result in

$$\mathbf{P}^h = \mu \mathbf{F}^h + \frac{\lambda}{2} J^h \tilde{\mathbf{F}}^{-T} - (\frac{\lambda}{2}+\mu)\frac{1}{J^h}\tilde{\mathbf{F}}^{-T} \tag{20}$$

With the help of the above, balance of linear momentum (1) is approximated as

$$\mathbf{R}^h = \text{Div}(\mathbf{P}^h) + \rho_0 \mathbf{b} - \rho_0 \dot{\mathbf{v}}^h \tag{21}$$

where $\dot{\mathbf{v}}^h = \ddot{\mathbf{x}}^h$. Note, generally speaking, $\mathbf{R}^h$ does not vanish identically since kinematic approximation is implied. It is a function of $n^{dof}$ IDFs $b_p(t)$, $(p=1,..,n^{dof})$. Next, we adopt a weak formulation to derive $(n^{shf}+1)$ vector equations of motion and consequently $n^{dof}$ scalar ODEs.

## 2.2 Equations of motion.

A weak formulation is used to restrict the residual $\mathbf{R}^h$ in the domain, and to obtain the governing dynamical equations of motion of the system

$$\mathbf{R}_i(R,\gamma_m;E,\nu,\rho_0;b_p(t),\ddot{b}_q(t)) = \int_{\Omega_0} \mathbf{R}^h N_i dV = \mathbf{0} \tag{22}$$

$$(m=1,2,3), (i=0,..,n^{shf}), (p,q=1,..,n^{dof})$$

The above weak formulation defines $(n^{shf}+1)$ vector coupled ordinary differential equations $\mathbf{R}_i$. With the help of partial integration, divergence theorem and introduction of the traction boundary condition $\bar{\mathbf{t}} = \mathbf{P}\mathbf{N}$ ($\mathbf{N}$ is normal to the boundary in initial configuration), (22) is given by (e.g. [35] pp.84)

$$\int_{\Omega_0} N_i \rho_0 \dot{\mathbf{v}}^h dV = \int_{\Omega_0} N_i \rho_0 \mathbf{b} dV + \int_{\partial\Omega_0} N_i \bar{\mathbf{t}} dA - \int_{\Omega_0} \mathbf{P}^h \text{Grad}(N_i) dV \ , \ (i=0,..,n^{shf}) \tag{23}$$

where $\text{Grad}(N_i) = \sum_{k=1}^{3} N_{i,k} \mathbf{G}^k$. Next, definitions for mass coefficients - $\hat{M}_{ij}$, internal forces - $\mathbf{f}_i^{int}$, external and body forces - $\mathbf{f}_i^{ext}, \mathbf{f}_i^{body}$, are introduced

$$\mathbf{f}_i^{ext} = \int_{\partial\Omega_0} N_i \bar{\mathbf{t}} dA$$

$$\mathbf{f}_i^{body} = \int_{\Omega_0} N_i \rho_0 \mathbf{b} dV \ , \ (i=0,..,n^{shf}) \tag{24}$$

$$\hat{M}_{ij} = \rho_0 \int_{\Omega_0} N_i N_j dV$$

$$\mathbf{f}_i^{int} = \int_{\Omega_0} \mathbf{P}^h \text{Grad}(N_i) dV \ , \ (i,j=0,..,n^{shf}) \tag{25}$$

With the help of (24) and (25), the (vector) equations of motion (23), are written as follows

$$\mathbf{R}_i = \sum_{j=0}^{n^{shf}} \hat{M}_{ij} \ddot{\mathbf{u}}_j + \mathbf{f}_i^{int} - \mathbf{f}_i^{ext} - \mathbf{f}_i^{body} = \mathbf{0} \ , \ (i=0,..,n^{shf}) \tag{26}$$

Importantly, it is possible to write explicit closed-forms of equations $\mathbf{R}_i$ in terms of the initial geometry parameters $R, \gamma_1, \gamma_2, \gamma_3$, material constants $E, \nu, \rho_0$, and scalar functions $\ddot{b}_p(t), b_q(t)$, $(p,q=1,..,n^{dof})$, with no need in numerical integration for internal forces. To this end, we consider $\mathbf{P}^h$ given by (20), systematic approximation by means of Taylor's multivariable expansion about the middle point $\mathbf{X}_0$ is applied, the minimum order of expansion which ensures consistency with linear elasticity is $(\alpha-1)$ see [5]. Using shape functions (7) and operator (9) expansion is given by

$$\mathbf{P}^h \approx \mathbf{P}^h_{(\beta)} = \sum_{j=0}^{n^\beta} N_j \mathbf{P}_j(t) \ , \ \beta = \alpha - 1$$

$$\alpha = 1 \Rightarrow \beta = 0 \ , \ n^\beta = 0 \ , \ \alpha = 2 \Rightarrow \beta = 1 \ , \ n^\beta = 3$$
$$\alpha = 3 \Rightarrow \beta = 2 \ , \ n^\beta = 9 \ , \ \alpha = 4 \Rightarrow \beta = 3 \ , \ n^\beta = 19 \quad (27)$$
$$\alpha = 5 \Rightarrow \beta = 4 \ , \ n^\beta = 34$$

$$\mathbf{P}_0 = \mathbf{P}^h\Big|_{\mathbf{X}=\mathbf{X}_0} \ , \ \mathbf{P}_1 = \frac{\partial \mathbf{P}^h}{\partial \hat{X}_1}\Big|_{\mathbf{X}=\mathbf{X}_0} \ , \ \ldots \ , \ \mathbf{P}_9 = \frac{\partial^2 \mathbf{P}^h}{2 \partial \hat{X}_3 \partial \hat{X}_3}\Big|_{\mathbf{X}=\mathbf{X}_0} \ , \ \ldots$$

Using the above, together with (4), the integrand of internal forces (25) becomes

$$\mathbf{P}^h \mathrm{Grad}(N_i) \approx \sum_{j=0}^{n^\beta} N_j \mathbf{P}_j(t) \sum_{k=1}^{3} N_{i,k} \mathbf{G}^k = \sum_{j=0}^{n^\beta} \sum_{k=1}^{3} \mathbf{P}_j(t) \mathbf{G}^k N_j N_{i,k}$$

$$dV = \underbrace{G^{1/2}}_{1} d\hat{X}_1 d\hat{X}_2 d\hat{X}_3 = d\hat{X}_1 d\hat{X}_2 d\hat{X}_3 = r^2 \sin(\phi) dr d\phi d\theta \quad (28)$$

Lastly, the integrand (28) and the integrand of mass coefficients (25) are computed and then transformed to spherical coordinates (4), consequently, internal forces and mass coefficients are computed as

$$\hat{M}_{ij} = \rho_0 \int_{\frac{(\pi-\gamma_3)}{2}}^{\frac{(\pi+\gamma_3)}{2}} \int_{\frac{(\pi-\gamma_2)}{2}}^{\frac{(\pi+\gamma_2)}{2}} \int_{R-\gamma_1}^{R} N_i N_j r^2 \sin(\phi) dr d\phi d\theta$$

$$\mathbf{f}_i^{int} = \sum_{j=0}^{n^\beta} \sum_{k=1}^{3} \mathbf{P}_j(t) \mathbf{G}^k \int_{\Omega_0} N_j N_{i,k} r^2 \sin(\phi) dr d\phi d\theta \ , \ (i,j = 0,..,n^{shf}) \quad (29)$$

Throughout the study, symbolic algebra package-Maple$^{TM}$ were extensively used to compute all the explicit expressions. Moreover, specific case of thick shell-hemisphere Fig. 1(c) was considered.

## 3. Linearization and formulation of the free vibration problem

In the previous section, vector equation of motion has been derived. Herein, derivation of a set of $n^{dof}$ scalar equations of motion is presented, linearization is carried out, mass and stiffness matrices are defined, and an eigenvalue problem is formulated for the free vibration analysis. To this end, we define

$$F_{3i+k}\left(R, \gamma_m; E, \nu, \rho_0; \ddot{b}_p(t), b_q(t)\right) = \mathbf{R}_i \cdot \mathbf{e}_k = 0$$
$$(i = 0,..,n^{shf}), (m,k = 1,2,3), (p,q = 1,..,n^{dof}) \tag{30}$$

and

$$F_{3i+k}^{mass}(R, \gamma_m, \rho_0; \ddot{b}_p(t)) = \sum_{j=0}^{n^{shf}} \hat{M}_{ij} \ddot{\mathbf{u}}_j \cdot \mathbf{e}_k$$

$$F_{3i+k}^{stiff}(R, \gamma_m, E, \nu; b_p(t)) = \mathbf{f}_i^{int} \cdot \mathbf{e}_k \tag{31}$$

$$\left(i = 0,..,n^{shf}\right), (m,k = 1,2,3), \left(p = 1,..,n^{dof}\right)$$

Free-vibration problem of a hemisphere is investigated, therefore external and body forces are neglected. These enable us to compactly rewrite (30) as

$$F_p = F_p^{mass} + F_p^{stiff} = 0, \left(p = 1,..,n^{dof}\right). \tag{32}$$

Next, we introduce the following notations

$$[\ddot{b}] = \begin{bmatrix} \ddot{b}^1 \\ . \\ . \\ . \\ \ddot{b}^{n^{dof}} \end{bmatrix}, [b] = \begin{bmatrix} b^1 \\ . \\ . \\ . \\ b^{n^{dof}} \end{bmatrix}, [F] = [F^{mass}] + [F^{stiff}] = \begin{bmatrix} F_1^{mass} \\ . \\ . \\ F_{n^{dof}}^{mass} \end{bmatrix} + \begin{bmatrix} F_1^{stiff} \\ . \\ . \\ F_{n^{dof}}^{stiff} \end{bmatrix} = [0] \tag{33}$$

Linearization of the system $[F]=[0]$ is carried out about the initial configuration, namely $[b]=[0]$, such that $[F] \approx [F]_{[b]=[0]} + \dfrac{\partial[F]}{\partial[b]}\bigg|_{[b]=[0]}[b]$. Using definitions(31), notation(33), the mass and stiffness matrices of the system of ODE (32) can written as:

$$[K] = \begin{bmatrix} K_{11} & . & K_{1n^{dof}} \\ . & . & . \\ K_{n^{dof}1} & . & K_{n^{dof}n^{dof}} \end{bmatrix}, K_{ij} = \frac{\partial F_i^{stiff}}{\partial b_j}\bigg|_{[b]=[0]}$$

$$[M] = \begin{bmatrix} M_{11} & . & M_{1n^{dof}} \\ . & . & . \\ M_{n^{dof}1} & . & M_{n^{dof}n^{dof}} \end{bmatrix}, M_{ij} = \frac{\partial F_i^{mass}}{\partial \ddot{b}_j} \tag{34}$$

Both matrices are symmetric. In addition, the mass matrix is positive definite and the stiffness matrix is positive semi-definite due to rigid body motion modes. Using the above definition, the linearization of the system (32) about the initial configuration $[b]=[0]$ is given by

$$[F] \approx [M][\ddot{b}] + [K][b] = [0] \tag{35}$$

Next it is assumed that $[b] = [\tilde{b}]\sin(\omega t)$ were $\omega$ is the vibration natural frequency and $[\tilde{b}]$ is the mode of the vibration (constant algebraic vector). The second time derivative becomes $\left[\ddot{\tilde{b}}\right] = -\omega^2 \left[\tilde{b}\right] \sin(\omega t)$, and after substituting to (35) the well-known form for small vibration problem in many degree of freedom (MDOF) system is

$$\left(-\omega^2 [M] + [K]\right)\left[\tilde{b}\right]\sin(\omega t) = [0] \tag{36}$$

Non-trivial solutions for $\left[\tilde{b}\right]$ exists if and only if $\left|-\omega^2 [M] + [K]\right| = 0$ (determinant vanishes). Solution of this problem leads to $n^{dof}$ natural modes and frequencies. The lowest six natural frequencies are zero, and represent rigid body translation and rotation.

### 4. Natural frequencies – free vibration 3D hemisphere:

The main objective of the below chapter is to explicitly show that the above formulation can be used to produce new, simple, closed-form results with significant engineering importance. Comprehensive closed-form study of small-vibration problem is out of scope of current letter. Herein, in order to demonstrate the ability to capture truly 3D modes, thick shell is taken to one of its extremes-hemisphere see Fig1(c).

All the expressions for resonant frequencies for all orders which has been addressed ($1^{st}$-$5^{st}$) throughout the research take the next form

$$\omega_k^2 = \frac{E}{\rho_0 R^2} \bar{\omega}_k^2(\nu) \ , \ (k = 1, 2, 3...) \tag{37}$$

where $\bar{\omega}_k$ is a non-dimensional frequency. This result provides an essential practical insight regarding the role of material and geometrical properties of a free vibrating body. Finally, we note that although some of the below expressions involve complex terms, *all frequencies are real*. This is a direct consequence of the fact that the mass and stiffness matrices are positive definite and semi-positive definite, respectively, in addition to both being real and symmetric (e.g.[38] pp. 293).

**First order approximation:** $\alpha = 1$, $n^{shf} = 3, n^{dof} = 12$. We begin by calculating the natural frequencies associated with the first order approximation, resulting in homogeneous deformation gradient $\mathbf{F}^h = \mathbf{F}^h(t)$ (see (17),(18)). To this end, the stiffness and mass matrices (34) are formulated and the eigenvalue problem (36) is solved. The number of natural frequencies is equal to the size of matrices (34) namely $n^{dof}$. Also, as stated earlier, the first 6 natural frequencies match rigid body motion modes (three rigid body translation and three rotation modes) are zero. The remaining 6 non-trivial natural frequencies are

$$\bar{\omega}_1^2 = \bar{\omega}_2^2 = \frac{5}{(1+\nu)}$$

$$\bar{\omega}_3^2 = \bar{\omega}_4^2 = \frac{415}{38(1+\nu)}$$

$$\bar{\omega}_5^2 = \frac{-415 + 320\nu + 15\sqrt{225 - 640\nu + 1536\nu^2}}{38(2\nu^2 - 1 + \nu)}$$

$$\bar{\omega}_6^2 = \frac{-415 + 320\nu - 15\sqrt{225 - 640\nu + 1536\nu^2}}{38(2\nu^2 - 1 + \nu)}$$

(38)

All the above eigenvalues (squared normalized natural frequencies) are associated with homogeneous deformation modes (eigenvectors). In (38) there are two frequencies which are repeated twice. Linear combinations of the modes that belong to the same eigenvalue are eigenvectors as well. Next, we excel this solution by providing better accuracy and richer spectral analysis by means of higher order approximations.

**Second order approximation** $\alpha = 2$, $n^{shf} = 9$, $n^{dof} = 30$. The second order approximation has $30 - 6 = 24$ non-trivial frequencies, four times more than the previous solution (which provided 6 nontrivial expressions (four of them are distinct). Formulating (34) and solving the corresponding eigenvalue problem (36), equivalently, one solves characteristic polynomial of rank $n^{dof}$ which is reduced to $n^{dof} - 6$ due to rigid body modes (zeroes). Unfortunately, deriving closed form solutions of polynomial with rank higher than 4 is not necessarily possible, hence, some natural frequencies are given explicitly and other implicitly (as roots of polynomial). The explicit squared normalized resonant frequencies are given by

$$\bar{\omega}_1^2 = \bar{\omega}_2^2 = \frac{14}{1+\nu} \; , \; \bar{\omega}_3^2 = \frac{11.061}{1+\nu} \; , \; \bar{\omega}_4^2 = \bar{\omega}_5^2 = \frac{23.016}{1+\nu}$$

$$\bar{\omega}_6^2 = \bar{\omega}_7^2 = \bar{\omega}_{fundamental}^2 = \frac{1.96642}{(1+\nu)} \; , \; \bar{\omega}_8^2 = \bar{\omega}_9^2 = \frac{5.07906}{(1+\nu)}$$

(39)

Next, we provide two polynomials of rank 5. The second polynomial is repeated twice, consequently, each of his roots (squared normalized natural frequencies) has multiplicity equal 2.

$$(486 - 486\nu + 3888\nu^4 + 1944\nu^5 - 2430\nu^2 + 486\nu^3)Z^5 + (-134742\nu^4 + 67371\nu +$$
$$+183069\nu^2 - 80067\nu^3 - 61023)Z^4 + (2373684\nu^3 - 2909508\nu - 2876940\nu^2 +$$
$$+2341116)Z^3 + (-33402600 + 40528740\nu - 2056880\nu^2)Z^2 + (-113582000\nu +$$
$$+166462800)Z - 247646000 = 0 \; , \; Z = \bar{\omega}^2(\nu)$$

$$(-972 - 1944\nu + 6804\nu^4 + 1944\nu^5 + 1944\nu^2 + 7776\nu^3)Z^5 + (-341655\nu^3 -$$
$$-146406\nu^4 - 146529\nu^2 + 146283\nu + 97563)Z^4 + (3129189\nu^2 + 3313296\nu^3 -$$
$$-3497403 - 3681510\nu)Z^3 + (54777247 + 35241633\nu - 19535614\nu^2)Z^2 +$$
$$+(-359422840 - 103369910\nu)Z + 757275400 = 0 \; , \; Z = \bar{\omega}^2(\nu)$$

(40)

Fig. 2 shows the dependence of the squares of natural frequencies on poison's ratio - $\nu$, for the second order solution. The implicit frequencies (40) are illustrated by the black lines, the explicit expressions (39) are in gold and the fundamental frequency (the frequency which is the lowest among all, in the positive $\nu$ range) given in red.

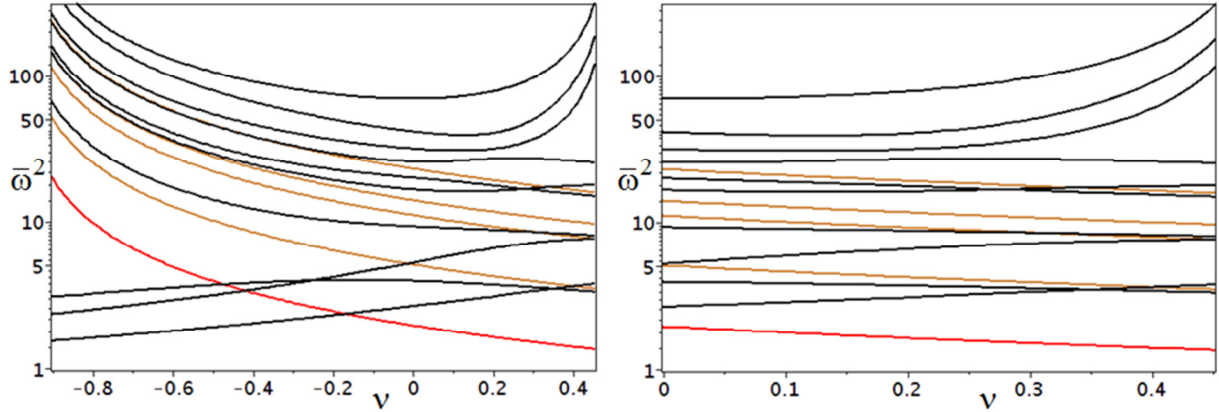

Figure 2: Second order solution showing square non-trivial natural frequencies of a free vibrating hemisphere as a function of Poisson's ratio. Also, frequencies which are given implicitly illustrated in black, explicit frequencies are in gold and explicit fundamental frequency denoted by red. The right plot is identical to the left, but focuses on the engineering range of $\nu > 0$.

Next, we examine the accuracy of the approximated solutions and demonstrate convergence. To this end, we compare between the 24 lowest frequencies (excluding the six trivial modes associated with rigid body motion) calculated by first, second, third, fourth and fifth order approximations to the results of a finite elements method (FEM). The finite elements analysis was performed with the commercial software ABAQUS$^{TM}$ 6.10, using 92595 standard 10 node full integration tetrahedral element *C3D10*, and for the following typical parameters $E = 210[GPa]$, $\rho_0 = 7800[Kg/m^3]$, $\nu = 0.3$, $R = 1[m]$. However, Table 1 has been normalized for convenience (37) consequently consists of $\bar{\omega}$ values.

| Mode # | FEM | $\alpha = 1$ | $\alpha = 2$ | $\alpha = 3$ | $\alpha = 4$ | $\alpha = 5$ | Err % $\alpha = 1$ | Err % $\alpha = 2$ | Err % $\alpha = 3$ | Err % $\alpha = 4$ | Err % $\alpha = 5$ |
|---|---|---|---|---|---|---|---|---|---|---|---|
| 1 | 1.1 | 2.0 | 1.2 | 1.1 | 1.1 | 1.1 | 84.6 | 15.7 | 3.2 | 0.3 | 0.1 |
| 2 | 1.1 | 2.0 | 1.2 | 1.1 | 1.1 | 1.1 | 84.6 | 15.7 | 3.2 | 0.3 | 0.1 |
| 3 | 1.6 | 2.5 | 1.8 | 1.7 | 1.6 | 1.6 | 54.9 | 13.9 | 6.3 | 0.6 | 0.2 |
| 4 | 1.7 | 2.9 | 1.9 | 1.8 | 1.7 | 1.7 | 67.2 | 9.0 | 1.0 | 0.4 | 0.0 |
| 5 | 1.7 | 2.9 | 1.9 | 1.8 | 1.7 | 1.7 | 67.2 | 9.0 | 1.0 | 0.4 | 0.0 |
| 6 | 1.7 | 5.1 | 2.0 | 1.9 | 1.8 | 1.8 | 192.9 | 13.4 | 7.8 | 1.2 | 0.7 |
| 7 | 1.7 | --- | 2.0 | 1.9 | 1.8 | 1.8 | --- | 13.4 | 7.8 | 1.2 | 0.7 |
| 8 | 1.8 | --- | 2.7 | 2.1 | 1.9 | 1.8 | --- | 51.5 | 21.7 | 5.2 | 0.0 |
| 9 | 1.8 | --- | 2.9 | 2.1 | 1.9 | 1.8 | --- | 65.0 | 21.7 | 5.2 | 0.0 |
| 10 | 2.2 | --- | 2.9 | 2.4 | 2.3 | 2.2 | --- | 32.4 | 7.1 | 4.7 | 0.8 |
| 11 | 2.2 | --- | 2.9 | 2.4 | 2.3 | 2.2 | --- | 32.6 | 9.4 | 5.0 | 0.8 |

| | | | | | | | | | | |
|---|---|---|---|---|---|---|---|---|---|---|
| 12 | 2.3 | --- | 3.3 | 2.4 | 2.3 | 2.3 | --- | 43.0 | 4.8 | 0.6 | 0.0 |
| 13 | 2.4 | --- | 3.3 | 2.6 | 2.4 | 2.4 | --- | 37.9 | 8.7 | 2.3 | 0.8 |
| 14 | 2.4 | --- | 4.1 | 3.0 | 2.7 | 2.6 | --- | 70.7 | 27.2 | 11.6 | 9.2 |
| 15 | 2.4 | --- | 4.1 | 3.0 | 2.7 | 2.6 | --- | 69.4 | 26.2 | 10.8 | 8.4 |
| 16 | 2.6 | --- | 4.1 | 3.1 | 2.9 | 2.6 | --- | 59.0 | 18.9 | 12.0 | 1.5 |
| 17 | 2.6 | --- | 4.2 | 3.1 | 2.9 | 2.6 | --- | 62.8 | 18.9 | 12.0 | 1.5 |
| 18 | 2.7 | --- | 4.2 | 3.3 | 3.0 | 2.7 | --- | 54.4 | 20.3 | 10.0 | 0.2 |
| 19 | 2.7 | --- | 5.2 | 3.3 | 3.0 | 2.7 | --- | 90.8 | 20.3 | 10.0 | 0.2 |
| 20 | 2.8 | --- | 5.2 | 3.8 | 3.0 | 2.8 | --- | 89.0 | 39.3 | 10.5 | 0.7 |
| 21 | 2.8 | --- | 6.1 | 3.8 | 3.0 | 2.8 | --- | 120.5 | 39.3 | 10.5 | 0.7 |
| 22 | 2.9 | --- | 7.1 | 3.9 | 3.2 | 3.1 | --- | 144.8 | 33.4 | 10.7 | 6.3 |
| 23 | 2.9 | --- | 7.1 | 4.2 | 3.2 | 3.1 | --- | 144.8 | 45.6 | 10.7 | 6.3 |
| 24 | 3.0 | --- | 9.9 | 4.2 | 3.2 | 3.1 | --- | 229.5 | 41.2 | 7.8 | 4.4 |

Table 1: Comparison between the 24 lowest natural frequencies calculated by $1^{st}$, $2^{nd}$, $3^{nd}$, $4^{th}$, $5^{th}$ order approximations and by finite elements method (FEM). Error is calculated with respect to FEM results.

The results of the comparison between the FEM and our approximated solution are outlined in Table 1. It is evident that higher order approximations enhance the accuracy of the solution, as expected. For each mode/row in the table, a higher order leads to increased accuracy. The values associated with the red line in Fig. 2, denoted by red.

**Fundamental frequency $\alpha = 2$.** The lowest resonant frequency is called fundamental frequency, for most engineering cases it is an important one (e.g.[38] pp.303). Fig. 2 shows that for all practical range $v > 0$, our formulation based on second order approximation $\alpha = 2$, resulted in simple, closed-form, explicit expression for fundamental frequency (see (37),(39)) with estimated accuracy (see Table 1) of about 16%

$$\omega^2_{fundamental} = \frac{E}{\rho_0 R^2} \frac{1.96642}{(1+v)} \tag{41}$$

Importantly, we have applied the derived shell mode to the extreme case of thick shell-hemisphere, and succeeded to derive new simple closed form result for 3D free vibrating body.